\documentclass[pmlr]{jmlr}

\usepackage{longtable}
\usepackage{multirow}
 \usepackage{bm}
 \usepackage{lscape}

 \usepackage{booktabs}
 
\usepackage[load-configurations=version-1]{siunitx} 

\makeatletter
\def\set@curr@file#1{\def\@curr@file{#1}} 
\makeatother

\theorembodyfont{\upshape}
\theoremheaderfont{\scshape}
\theorempostheader{:}
\theoremsep{\newline}

\jmlrvolume{219}
\jmlryear{2023}
\jmlrworkshop{Machine Learning for Healthcare}

\title[Graph representation learning from EHRs and familial relationships]{Characterizing personalized effects of family information on disease risk using graph representation learning}

\author{\Name{Sophie Wharrie}
       \Email{sophie.wharrie@aalto.fi}\\ 
       \addr Department of Computer Science, Aalto University, Finland\\
       \AND
       \Name{Zhiyu Yang}
       \Email{zhiyu.yang@helsinki.fi}\\ 
       \addr Institute for Molecular Medicine Finland, FIMM, University of Helsinki, Finland\\
       \AND
       \Name{Andrea Ganna}
       \Email{aganna@broadinstitute.org}\\ 
       \addr Institute for Molecular Medicine Finland, FIMM, University of Helsinki, Finland; Broad Institute of MIT and Harvard, Cambridge, MA, USA; Massachusetts General Hospital, Cambridge, MA, USA\\
       \AND
       \Name{Samuel Kaski}
       \Email{samuel.kaski@aalto.fi}\\ 
       \addr Department of Computer Science, Aalto University, Finland; Department of Computer Science, University of Manchester, UK
       }

\begin{document}

\maketitle

\begin{abstract}
  Family history is considered a risk factor for many diseases because it implicitly captures shared genetic, environmental and lifestyle factors. Finland's nationwide electronic health record (EHR) system spanning multiple generations presents new opportunities for studying a connected network of medical histories for entire families. In this work we present a graph-based deep learning approach for learning explainable, supervised representations of how each family member's longitudinal medical history influences a patient's disease risk. We demonstrate that this approach is beneficial for predicting 10-year disease onset for 5 complex disease phenotypes, compared to clinically-inspired and deep learning baselines for Finland's nationwide EHR system comprising 7 million individuals with up to third-degree relatives. Through the use of graph explainability techniques, we illustrate that a graph-based approach enables more personalized modeling of family information and disease risk by identifying important relatives and features for prediction.
\end{abstract}

\section{Introduction}

Family history is a well-established indicator of a patient's predisposition to certain health risks. However, for common chronic diseases with complex etiologies, such as coronary artery disease and type 2 diabetes, assessment of family history is complicated by the interactions of multiple genetic, environmental and lifestyle factors \citep{wilson2009systematic, valdez2010family}. 

The wider availability of electronic health record (EHR) systems in recent years has made it possible to develop deep learning approaches for studying a patient's detailed medical history. Beyond a single patient EHR, nationwide and regional-level EHR systems spanning multiple generations present a new opportunity for studying connected networks of EHRs for entire families. Our in-house dataset comprising three decades of nationwide EHR data for over 7 million individuals in Finland enables a new kind of deep learning study for up to third-degree relatives.

Previous works recognize that it is beneficial to include features for family history in machine learning models that use clinical and biological data for predicting the risk of various diseases \citep{behravan2020predicting, kinreich2021predicting}. However, machine learning approaches developed for tabular data do not explicitly model the underlying geometric structure of family history that is in itself informative for disease prediction. Specifically, the proximity of a genetic relationship, number of relatives affected by a disease or (potentially many) associated risk factors, and relatives' age of diagnosis for these conditions, are relevant details that we hypothesize are more appropriately modeled using geometric deep learning methods. Furthermore, EHRs contain thousands of potentially informative variables that require new machine learning tools to analyze at the scale of entire families.

In this work we consider recasting the prediction of disease risk incorporating family history from EHRs as a graph modeling problem. Graphs provide an expressive data structure for representing high-dimensional, longitudinal EHR data at the individual level for each family member, and the connectivity structure of genetic relatedness encodes prior knowledge about the importance of different types of family members for predicting an individual's disease risk. The main contributions of our work can be summarized as follows:
\begin{itemize}
    \item We introduce a scalable, disease-agnostic machine learning tool based on graph neural networks (GNNs) and long short-term memory networks (LSTMs) for learning supervised representations predicting a patient’s disease risk from family information;
    \item Through a study of nationwide EHR data for predicting 10-year disease onset of 5 complex diseases, we show that graph-based approaches better predict multifactorial diseases, compared to clinically-inspired and deep learning baselines for family history;
    \item Using graph explainability techniques we demonstrate that GNN-LSTM embeddings identify medical features of family members (from over 1500 features) that are more suitable for predicting disease risk than features identified by an epidemiological baseline.
\end{itemize}

We share the software\footnote{\url{https://github.com/dsgelab/family-EHR-graphs}} developed for this work and a synthetic dataset that models the key properties of the private dataset\footnote{\url{https://www.finregistry.fi/}}.

\subsection*{Generalizable Insights about Machine Learning in the Context of Healthcare}

Family history has long been considered a risk factor for many diseases, but in practice this information is often underutilized in health care due to a lack of suitable data collection and interpretation tools \citep{feero2008new, ginsburg2019family}. In this work we model EHRs as a connected network of medical histories for extended families and learn supervised graph representations for predicting the onset of disease. The generalizable machine learning insight we would like readers to take away from this work is that explicitly modeling the underlying geometric structure of family history provides an informative representation of the shared genetic, environmental and lifestyle factors within families that influence complex diseases. The software tools that we develop in this work conveniently scale to modeling thousands of variables for large family structures. We believe that the clinical relevance of this work lies in (1) demonstrating the value of disease history from family members beyond first-degree relatives and health history apart from the disease of interest in an individual's risk prediction; and (2) presenting GNNs as a new approach to integrating extensive information with complex familial connections, which can be used to improve family history anamnesis in a clinical setting.

\section{Related Work}

\subsection{Family history and disease prediction}

Our work presents a new machine learning framework for modeling the role of family history in disease risk. Clinicians routinely query patients about family history to inform clinical decision making, based on prior knowledge, clinical guidelines and risk assessment tools \citep{wilson2009systematic, valdez2010family, ginsburg2019family}. Different types of statistical approaches have been developed for modeling family history and heritable effects of disease. When genetics data is available, polygenic risk scoring methods can be used to quantify the genetic risk of disease \citep{lewis2020polygenic, li2020electronic}. BLUP (best linear unbiased prediction) is a phenotype prediction approach widely used in animal breeding, which typically models pedigree information as random effects in a linear mixture model and constructs the prediction outcome along with other fixed-effect predictors \citep{wang2018expanding}. Due to lack of availability of extensive pedigree data, the application of this approach has been limited in human populations. Machine learning approaches have been applied to combine different types of clinical and biological data with tabular (non-graph) features for family history \citep{behravan2020predicting, kinreich2021predicting}. However, to the best of our knowledge, we are the first to utilize a patient's family history as a graph of EHRs for up to third-degree relatives, which we model using a geometric deep learning approach.

\subsection{Graph neural networks in healthcare}
In this work we model family information as a graph data structure using graph neural networks (GNNs). Graph representation learning methods such as GNNs have achieved leading performance for a number of machine learning tasks with graph-structured data in biomedicine and healthcare, including modeling protein-protein interactions, drug-drug interactions and disease associations \citep{li2022graph}. In a clinical context, GNNs have been applied to EHRs expressed as graph data structures, for example, connecting lab tests and medications associated with medical encounters \citep{mao2022medgcn, wu2019representation}. Other works have applied GNNs to EHR data for graphs constructed on the basis of patient similarity, where nodes represent individual patients and edges represent similarities between patients. For example, \cite{parisot2018disease} measure similarity in terms of demographic features and phenotypes derived from imaging data while \cite{tong2022predicting} consider similar diagnoses. We define similarity between patients in terms of family relationships and study a form of graph classification task, where each patient's family graph (up to third-degree relatives) is used to predict a binary disease outcome.

\section{Methods}

\begin{figure}[t]
  \centering 
  \includegraphics[width=0.9\textwidth]{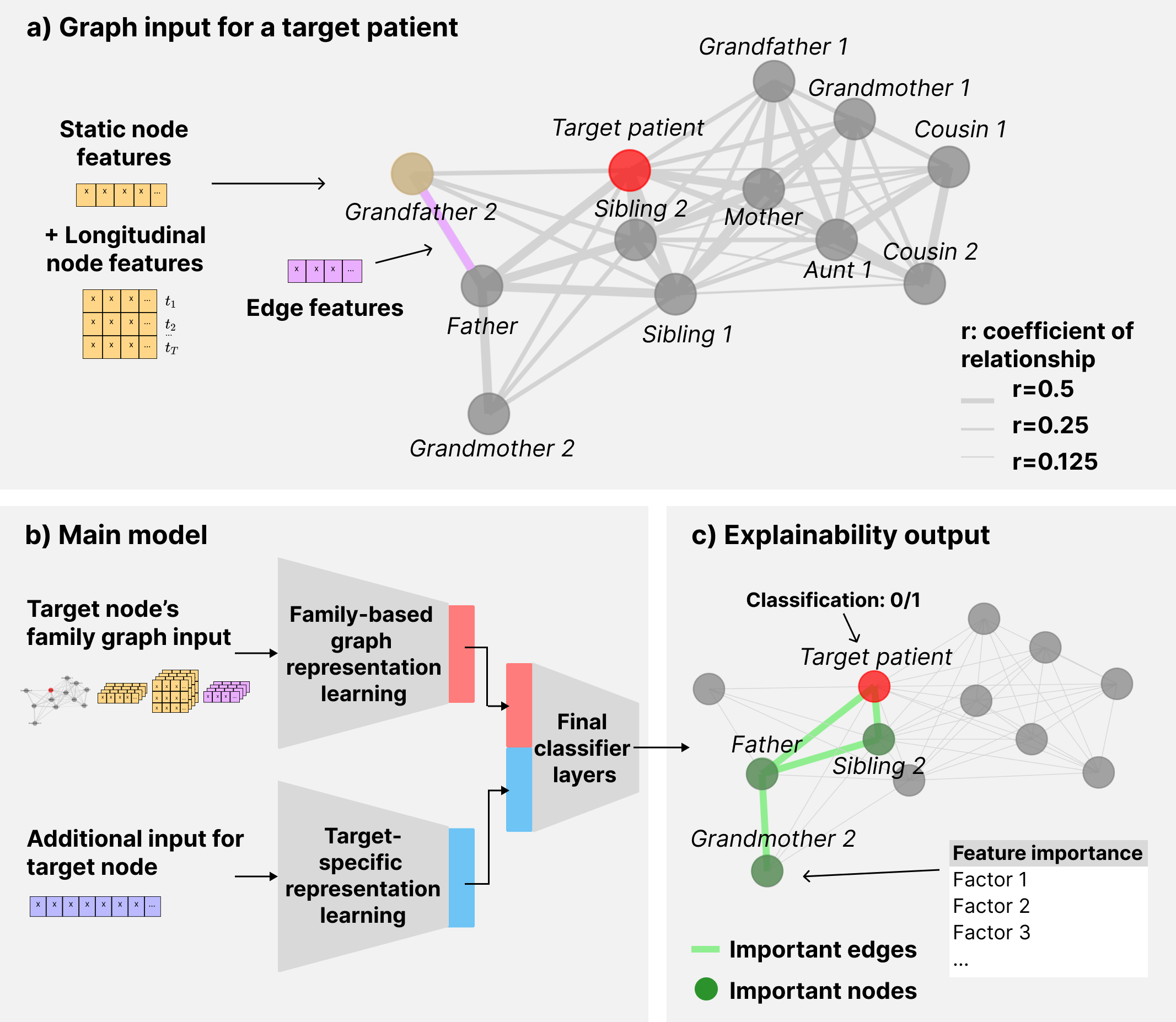} 
  \caption{Overview of modeling approach. a) Graph inputs contain the EHRs for (up to third-degree) family members as node features and descriptors of genetic relationships as edge features; b) The ML model jointly learns family (graph) and patient-specific representations, concatenated in the final layers to classify patient outcomes; c) Graph explainability analysis identifies important relatives (nodes) and risk factors/co-morbidities (node features) the model found important for predicting the patient’s risk.
   }
  \label{fig:graphical-abstract} 
\end{figure} 

An overview of the modeling approach developed in this work is given in Figure \ref{fig:graphical-abstract}. The following sections describe how we design a machine learning system that predicts a binary disease outcome for a patient and explains which relatives and medical features of those relatives were most important for the model in arriving at the prediction. 

\subsection{Preliminaries and notation}

To study the effects of family information on disease risk we consider the binary classification problem of predicting a future health outcome based on previously observed EHR data. We refer to the study cohort that we're predicting health outcomes for as the \textit{target samples}, denoted by the set $P_{target}$, and we also define a broader set of patients consisting of the target samples and their relatives, which we call the \textit{graph samples} and denote by $P_{graph}$. 

The data for target patients and their family members and the nature of the genetic relationships between them can be encoded using a graph data structure, as illustrated in Figure \ref{fig:graphical-abstract}a. A graph is typically defined as $\mathcal{G} := (V, E)$, where $V$ is a set of vertices (or nodes) and $E$ is a set of pairs of vertices called edges. In machine learning it is more convenient to define a graph as $\mathcal{G} := (\bm{X}, (\bm{I}, \bm{E}))$, where $\bm{X} \in \mathbb{R}^{N \times F}$ is a node-feature matrix of $F$ features for $N$ nodes and $(\bm{I}, \bm{E})$ is a tuple containing information about the $E$ edges, where $\bm{I} \in \mathbb{N}^{2 \times E}$ contains the indices of the edge pairs and $\bm{E} \in \mathbb{R}^{E \times D}$ is an edge-feature matrix of $D$ features for $E$ edges \citep{fey2019fast}. In this work we consider a dataset of the form $\mathcal{D} := \{(\mathcal{G}_i, \tilde{\bm{X}_i}, \bm{y}_i)\}_{i=1}^{|P_{target}|}$, where the aim is to predict a target patient's binary outcome $\bm{y}_i \in \{0, 1\}$ from their family graph $\mathcal{G}_i = (\bm{X}_i, (\bm{I}_i, \bm{E}_i))$ and own set of features $\tilde{\bm{X}_i}$.

\subsection{Constructing family graphs}

The graphs in geometric deep learning algorithms can be viewed as geometric priors that provide an inductive bias to the learning algorithm \citep{bronstein2021geometric}. Biologically speaking, first-degree relatives share more DNA than second- or third-degree relatives, and first-degree relatives are also more likely to share environmental conditions and lifestyle factors. In addition to the proximity of the genetic relationship, additional attributes of the family history such as the age of diagnosis of the relatives' condition and presence of associated risk factors are also informative when studying complex diseases \citep{valdez2010family}. 

Based on these assumptions, we construct a family graph $\mathcal{G}_i = (\bm{X}_i, (\bm{I}_i, \bm{E}_i))$ with nodes for the target patient $i \in P_{target}$ and their first-, second- and third-degree relatives in $P_{graph}$. For graph edges $\bm{I}_i$ representing genetic relatedness, we empirically compare three different configurations in synthetic data experiments (detailed results can be found in Figure \ref{fig:gnn-design-syn} of Appendix \ref{appendix:hyperparameter-choices-etc}): (1) parent/child edges (like a family tree or pedigree chart), (2) parent/child edges with additional edges linking each node directly to the target node, and (3) edges linking all genetically related nodes. This is motivated from a machine learning perspective, where graphs are often augmented using strategies such as virtual nodes and edges between two-hop neighbors to improve the effectiveness of the GNN \citep{10.1145/3535101}. Edge features $\bm{E}_i$ include the coefficient of relationship, $r$, and for GNN layers that support multiple edge features, we also add binary indicators describing the type of genetic relationship in relation to the target node (parent/child, sibling/grandparent etc). The node features consist of both static features (age, sex) and longitudinal features (diagnosis codes aggregated at the yearly level). An example graph is depicted in Figure \ref{fig:graphical-abstract}a and further details of the features used in experiments are given in Section \ref{sec:cohort}.

\subsection{Learning supervised graph representations}
\label{ssec:graphmethods}

\begin{figure}[t]
  \centering 
  \includegraphics[width=0.9\textwidth]{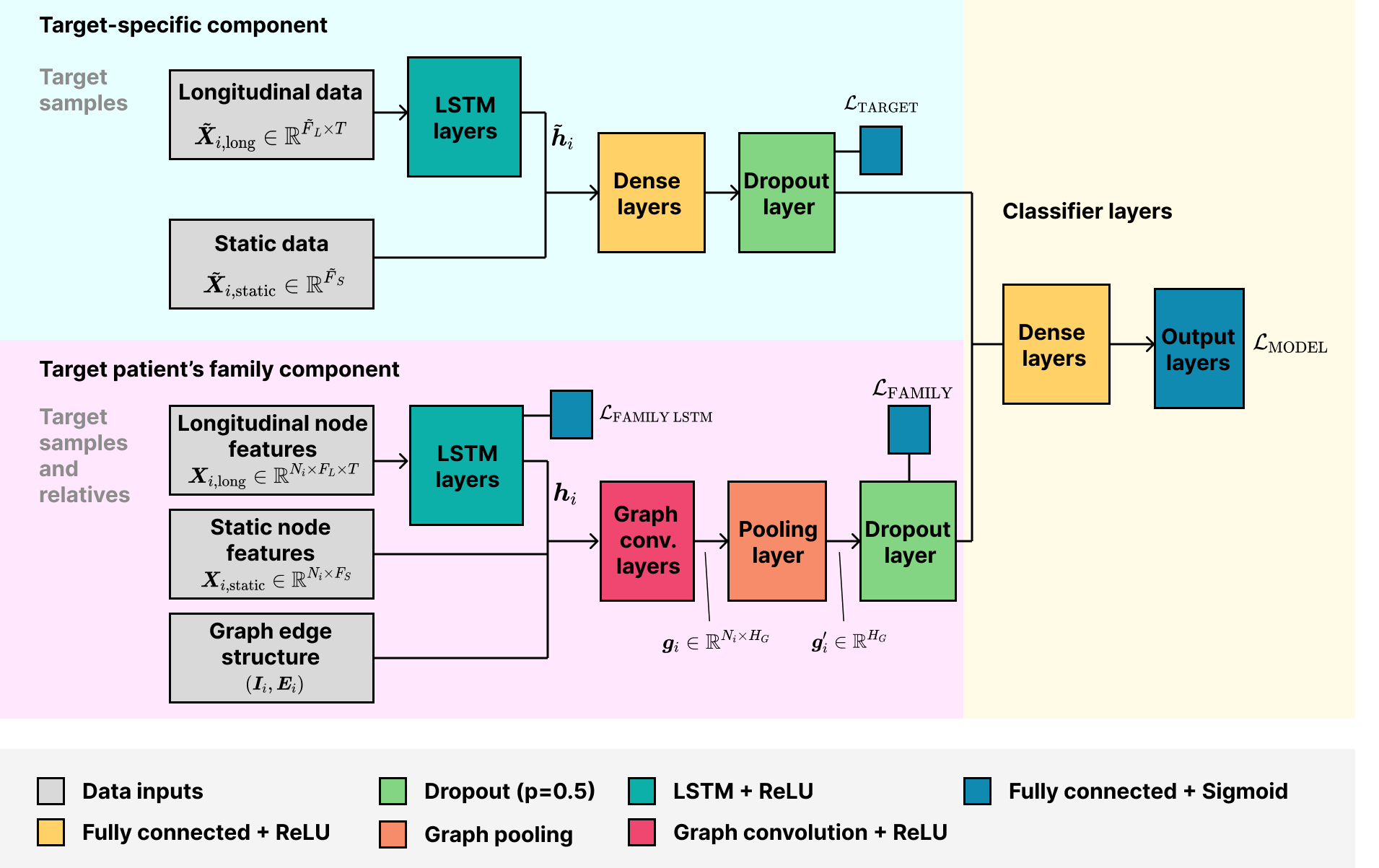} 
  \caption{Neural network architecture diagram. 
   }
  \label{fig:architecture-diagram} 
\end{figure} 

We develop a neural network architecture that learns to classify a patient's health outcome from both their own EHR data and a graph representation of their family's EHRs. Multiple input heads corresponding to the target individual's own data, $\tilde{\bm{X}}_i=(\tilde{\bm{X}}_{i,\text{static}},\tilde{\bm{X}}_{i,\text{long}})$, and their family's data, $\mathcal{G}_i=((\bm{X}_{i,\text{static}},\bm{X}_{i,\text{long}}), (\bm{I}_i,\bm{E}_i))$, jointly learn target- and family-based representations that are combined in the final classifier layers, as depicted in Figure \ref{fig:graphical-abstract}b. 

The full neural network architecture is illustrated in Figure \ref{fig:architecture-diagram}. The longitudinal inputs are reduced to $H_L$-dimensional embeddings, $\tilde{\bm{h}}_i \in \mathbb{R}^{H_L}$ for $\tilde{\bm{X}}_{i,\text{long}} \in \mathbb{R}^{\tilde{F}_L \times T}$ and $\bm{h}_i \in \mathbb{R}^{N_i \times H_L}$ for $\bm{X}_{i,\text{long}} \in \mathbb{R}^{N_i \times F_L \times T}$, using the final hidden state output of bidirectional LSTM layers; a frequently used architecture for longitudinal EHRs \citep{10.1145/3531326}. The component of the model corresponding to the target patient's data effectively corresponds to an LSTM: the embedding, $\tilde{\bm{h}}_i$, is concatenated with the static data, $\tilde{\bm{X}}_{i,\text{static}} \in \mathbb{R}^{\tilde{F}_S}$, followed by stacked dense layers and a dropout layer to prevent overfitting. The family component corresponding to the target samples and their relatives combines LSTM and GNN layers, similar to the approach proposed by \cite{tong2022predicting}. The graph convolutional layers learn an individual embedding for each node, $\bm{g}_i \in \mathbb{R}^{N_i \times H_G}$, with hidden dimension $H_G$, from $((\bm{h}_i, \bm{X}_{i,\text{static}}), (\bm{I}_i,\bm{E}_i))$.

The convolutional layers of a GNN act as propagation modules that aggregate information from neighboring nodes, followed by pooling layers that aggregate node-level representations into a graph-level representation \citep{zhou2020graph}. There are various types of propagation and pooling layers that have been developed for GNNs and therefore in experiments we evaluate the sensitivity of the model to various design choices (Figure \ref{fig:gnn-design-syn}). For example, a simple GCN layer updates a node embedding, $\bm{g}_{i,j}$, for family member $j$, at each layer, $l$, with the operation 

\begin{equation}
    \bm{g}_{i,j}^{(l+1)} = \sigma(\bm{\hat{D}}^{-\frac{1}{2}}\bm{\hat{A}}\bm{\hat{D}}^{-\frac{1}{2}} \bm{g}_{i,j}^{(l)} \bm{W}^{(l)}),
\end{equation}

\noindent where $\bm{\hat{D}}^{-\frac{1}{2}}\bm{\hat{A}}\bm{\hat{D}}^{-\frac{1}{2}}$ is a normalized adjacency matrix\footnote{$\bm{A}$ is an adjacency matrix of an undirected graph, $\bm{\hat{A}} = \bm{A} + \bm{I}$ is $\bm{A}$ with self-loops, and $\bm{\hat{D}}$ is a diagonal matrix containing the degree of each node}, $\bm{W}^{(l)}$ is the weight matrix of the neural network, and $\sigma$ is an activation function, such as $\text{ReLU}(\cdot) = \max(0,\cdot)$ \citep{kipf2016semi}. Intuitively, this amounts to taking a weighted sum of the embeddings for node $j$ and their neighboring nodes (relatives) in the graph. A pooling operation then obtains a graph-level embedding, $\bm{g}_i' \in \mathbb{R}^{H_G}$, from a set of node embeddings, $\bm{g}_i \in \mathbb{R}^{N_i \times H_G}$, by applying an operation such as a sum or average. 

The representations learned from the target and graph components of the neural network are combined in the final layers to predict the binary outcome $\hat{\bm{y}}_i$. \cite{tong2022predicting} observe that the quality of representation learning can degrade when LSTM layers are combined with a GNN. We adapt the solution proposed in their work and use a loss function of the form $$\mathcal{L} = \gamma \mathcal{L}_{\text{MODEL}} + \alpha \mathcal{L}_{\text{TARGET}} + \beta \mathcal{L}_{\text{FAMILY}} + \delta \mathcal{L}_{\text{FAMILY LSTM}},$$ with individual (weighted) binary cross-entropy (BCE) loss terms for the full model output, $\mathcal{L}_{\text{MODEL}}$, the target-specific representation learning component, $\mathcal{L}_{\text{TARGET}}$, the graph representation learning component, $\mathcal{L}_{\text{FAMILY}}$, and the LSTM embedding of the graph node features, $\mathcal{L}_{\text{FAMILY LSTM}}$, where $\gamma, \alpha, \beta, \delta \in \mathbb{R}$ are hyperparameters. Each individual loss term takes the form $$\mathcal{L}_j = -\frac{1}{N} \sum_{i=1}^N w_i [y_i \log(\hat{y}_{i,j}) + (1-y_i)\log(1-\hat{y}_{i,j})],$$ where $N=|P_{target}|$ is the number of target patients, $\hat{y}_{i,j}$ is obtained by applying fully connected and sigmoid layers where indicated by the corresponding $\mathcal{L}_j$ in Figure \ref{fig:architecture-diagram}, and $w_i=N/n_i$ is a class weight used because of class imbalance in the dataset, where $n_i$ is the number of samples in the same class as individual $i$. Our ablation studies evaluate the sensitivity of the model to various design choices, such as the use of family data and LSTMs (Table \ref{tab:ablation-fin}). Full details of the hyperparameter values used in experiments are given in Appendix \ref{appendix:hyperparameter-choices-etc} and we report computational times for the GNN-LSTM algorithm and baselines in Table \ref{tab:computational-time}.

\subsubsection{Additional considerations for model training}

In practice we need to consider characteristics of real medical data that create challenges for machine learning models, which for our dataset are primarily class imbalance and missing family data. Class imbalance arises because we are predicting diseases with prevalence $<10\%$ in the selected cohorts. We alleviate this by using class-weighted loss functions, sampling strategies designed for imbalanced classification, methods to prevent overfitting, and careful choice and interpretation of evaluation metrics. The specific details of these approaches are listed in Appendix \ref{app:class-imbalance}. Medical records for a target patient's family members may be missing because they fall outside of the coverage of the EHR system. In Section \ref{sec:cohort} we describe how we handle missing family data.

\subsection{Explainability approaches}
\label{sssec:explainability-methods}

The graph data structure and specifically the supervised node embeddings learned by the GNN-LSTM for each individual family member can be analyzed to explain the relatives and medical features of those relatives that the model found important for predicting a specific patient’s disease risk, as illustrated in Figure \ref{fig:graphical-abstract}c. We apply the model-agnostic, perturbatation-based explainability approach GNNExplainer \citep{amara2022graphframex, ying2019gnnexplainer} to estimate a soft mask of normalized node importance weights $\bm{M}_{imp, i} \in [0, 1]^{N_i}$ and node feature importance weights $\bm{X}_{imp, i} \in [0, 1]^{N_i \times F'}$ for each family graph\footnote{Graph explainability approaches are generally formulated in terms of edge importances, but for the application studied in this work it makes more sense to compute node importances}. GNNExplaner is applied as a post-training analysis and can be used to jointly learn node and node feature importances by maximizing the mutual information objective 

\begin{equation}
    \max_{\mathcal{G}_{i,S}, \bm{p}_i}{MI(\bm{y}_i, (\mathcal{G}_{i,S}, \bm{p}_i))} = H(\bm{y}_i) - H(\bm{y}_i \mid \mathcal{G}_i = \mathcal{G}_{i,S}, \bm{X}_{i,\text{long}} = \bm{X}_{i,\text{long},S}),
\end{equation}

\noindent where $\mathcal{G}_{i,S} \subset \mathcal{G}_i$ is a subgraph excluding a subset of nodes and $\bm{X}_{i,\text{long},S}$ is a subset of node features for the masked features $\bm{p}_i \in \{0,1\}^{F'}$. The mutual information $MI(\bm{y}_i, (\mathcal{G}_{i,S}, \bm{p}_i))$ quantifies the change in probability of the prediction of $\hat{\bm{y}_i}$ when the nodes and features are limited to $(\mathcal{G}_{i,S}, \bm{p}_i)$, with the idea that a large change in probability indicates that the excluded nodes and features were important for prediction. For the longitudinal input, $\bm{X}_{i,\text{long}} \in \mathbb{R}^{N \times F \times T}$, the feature mask is defined for an equivalent 2-dimensional input $\bm{X}_{i,\text{long}} \in \mathbb{R}^{N \times F'}$, where $F' = F \times T$. A node feature importance can be obtained by averaging $\bm{X}_{imp, i}$ across the time dimension. For data privacy reasons we cannot report individual-level patient data in our results, so global feature importances for all relatives of a certain type (parent, sibling, etc) are calculated by taking an average of node feature importances for all relatives of that type, weighted by node importances, $\bm{M}_{imp, i}$.

\section{Cohort}
\label{sec:cohort}

\begin{figure}[h]
  \centering 
  \includegraphics[width=0.8\textwidth]{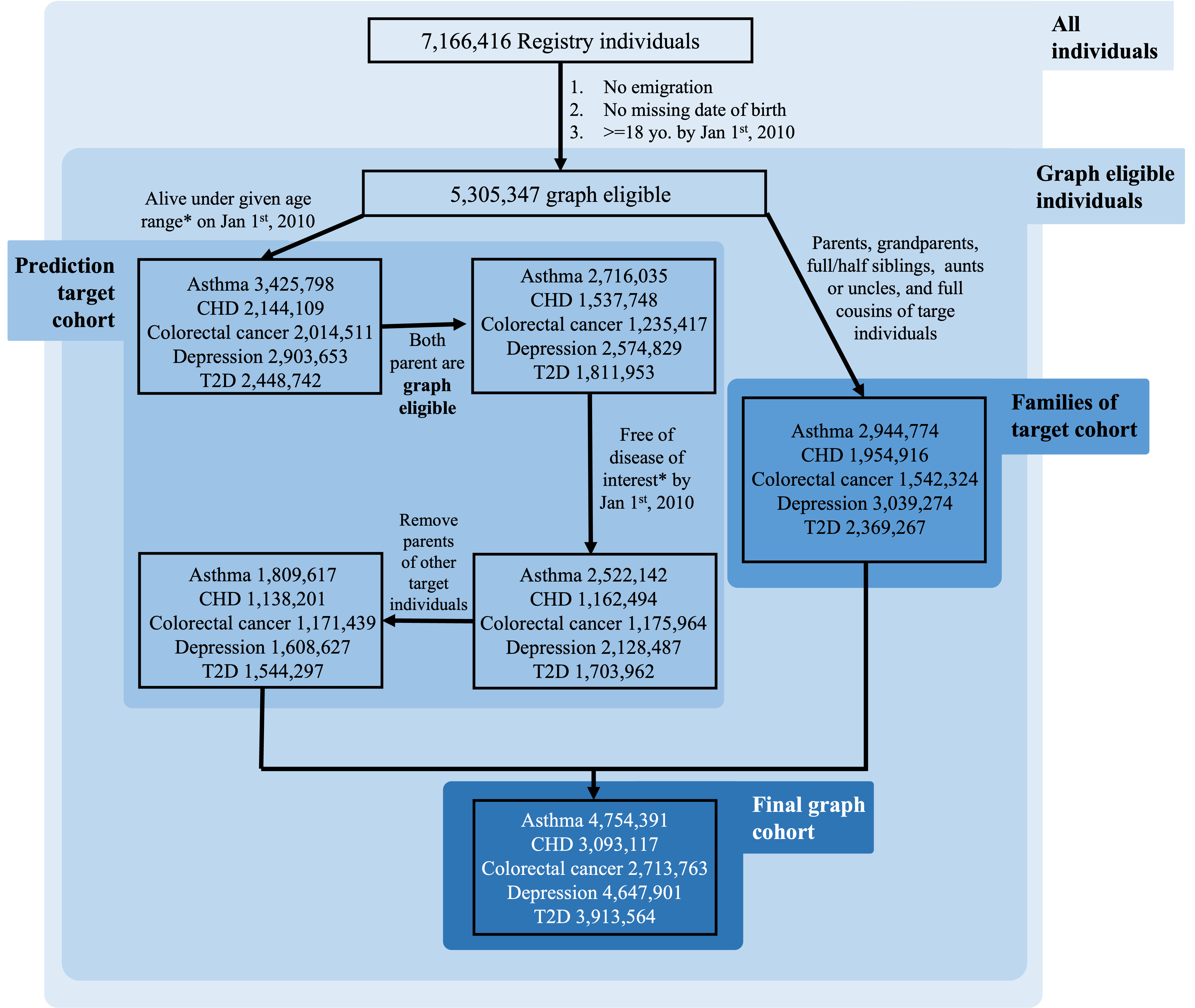} 
  \caption{Quality control and inclusion for each disease of interest. Also see Table \ref{tab:endpoint-definitions} for asterisk-noted disease-specific age ranges and control exclusion criteria.}
  \label{fig:cohort} 
\end{figure}

\subsection{Cohort Selection} 
The experiments are carried out using data from FinRegistry\footnote{\url{https://www.finregistry.fi/}}, which is an aggregated nationwide health registry dataset. Out of in total 7,166,416 available individuals, we excluded ones with emigration records, age less than 18 years old by Jan 1st, 2010, or missing sex or date of birth due to data incompleteness. We define the remaining 5,305,347 individuals as the graph eligible group, $P_{graph}$, and used them as the main cohort for most of the experiments. On top of that, for each selected disease, $d$, we further define a target cohort, $P_{target,d}$, for a disease prediction task as the subset of graph eligible individuals who satisfy the following criteria: (1) alive at a designated age on Jan 1st, 2010 (usually the age range suggested for disease screening); (2) both parents are also graph eligible individuals; (3) no medical record related to the disease of interest by Jan 1st, 2010; (4) not parents of any other individuals within the target cohort. See Figure \ref{fig:cohort} and Table \ref{tab:endpoint-definitions} for detailed disease specific cohort definitions. Once the target cohort is defined for each disease, we then extract graph eligible relatives of those individuals within the cohort to construct the family graph of each target patient. Family members included for each target individual are their parents, grandparents, full or half siblings, aunts or uncles, and cousins.

By our inclusion criteria, all individuals in the target cohort for each disease have both parents available in their subgraph, which makes the minimum size of a family graph for any target individual $N_i \geq 3$. Furthermore, all targets are free of the disease of interest by Jan 1st, 2010. Our machine learning task is to predict their disease onset during the window Jan 1st, 2011-Dec 31st, 2021, based on EHR data observed during Jan 1st, 1990-Jan 1st, 2010. Therefore, the target patients diagnosed with the disease of interest during the prediction window are defined as cases while the rest are defined as controls.

\subsection{Disease and Feature Choices} 
\label{ssec:features}
We select five common diseases as our prediction target: adult asthma, colorectal cancer, coronary heart disease, depression and suicide, and type two diabetes. These diseases were chosen due to their (1) high prevalence in the population; (2) high impact and Global Burden of Disease \citep{roth2018global}; (3) relatively high heritable impact and (4) differences in genetic architectures \citep{han2020genome,wray2018genome,nelson2017association,mahajan2018fine, law2019association}.
Model features consist of static features for age, sex and family history, and longitudinal EHR features. Family history features are constructed as categorical features for the aforementioned relatives up to Jan 1st, 2010, e.g., a target individual’s cousin’s disease history will be: 1 if any of their cousins has been diagnosed with the disease before Jan 1st, 2010, 0 otherwise. As mentioned above, all target individuals have disease history from both parents. For each of the other eligible relatives, we use an extra indicator variable to encode its availability since missingness of family members itself can be informative, i.e., on top of the cousin's disease history, an additional feature will have 1 for target individuals who have at least one cousin, and 0 otherwise. For the longitudinal EHR features, out of 3000 disease endpoints in our dataset\footnote{Details available at \url{https://risteys.finregistry.fi/}}, we select ones with occurrences $> 0.1\% \times$ sample size of the target cohort, resulting in $\sim 1500$ features. We consider only medical records within the period Jan 1st, 1990 - Jan 1st, 2010, which prevents label leakage across the overlap between target cohort and family members of targets. The endpoint data is aggregated at yearly intervals as binary features indicating if the endpoint was reported in that year.

\subsection{Synthetic dataset}
\label{ssec:cohort-synthetic}

Since we are using a private dataset, we also create and release synthetic pedigree and phenotype data for results reproducibility. Using the liability threshold model \citep{neale2005liability,holst2016liability}, we assume one's disease liability $L$ is constituted by several components as below:
$$ L = L_{Herr} + L_{Age} + L_{Sex} + \epsilon,$$
where $L_{Herr}$ is in a broad sense heritable risk covering any kind of familial impact, including genetic heritability and heritable environmental risk that can be obtained from parents; $L_{Age} = \beta_{age} \cdot Age$ and $L_{Sex} = \beta_{sex} \cdot Gender$ are effects from individuals' age and gender which can be generalized to any fixed-effect factors, and $\epsilon$ is individual specific noise on the phenotype. Individuals having liability beyond a certain cutoff threshold are assigned the disease label. Synthetic data is created by starting with an ancestry population with randomly assigned heritable risks, and assuming random mating to create the next generation. This is repeated until the target number of generations is reached. See Appendix \ref{AppxC} for full details of our synthetic data algorithm.

\section{Experiments} 

We conduct experiments for predictive performance, ablation studies and explainability using real patient data, and also conduct further ablation studies and analysis with synthetic datasets. Experiments comparing predictive performance use a training setup described in Appendix \ref{appendix:hyperparameter-choices-etc} and for the test dataset report the metrics AUC-ROC, AUC-PRC and Matthews correlation coefficient (MCC), as recommended for medical datasets with imbalanced classes \citep{hicks2022evaluation, saito2015precision}. For real data experiments, uncertainty estimates are calculated using the MC dropout approach \citep{gal2016dropout} to report the mean and 95\% confidence interval across 3 samples, while for synthetic data experiments the results are averaged across 5 random samples of synthetic datasets. 

\subsection{Comparison with clinical baselines}

Graph-based methods are compared to two clinical baselines for predicting disease onset $\bm{y}_i$, that approximate general recommendations for collecting family history in a clinical setting \citep{familymedicalhis, walker1990clinical, shaukat2021acg, expert2001executive, handelsman2015american}: 

\begin{itemize}
    \item \textbf{Rule-based clinical baseline (B1)}: rule-based approach where a patient is labeled high risk ($\hat{\bm{y}}_i=1$) if at least one (up to third-degree) relative has a history of the disease;
    \item \textbf{MLP-based clinical baseline (B2)}: multilayer perceptron (MLP) with 2 linear layers and static features for age, sex, and family history, as defined in Section \ref{ssec:features}.
\end{itemize}

\noindent The graph-based model is not directly comparable to baselines that do not use longitudinal data, so we also include in experiments a simpler version of the model without the longitudinal features. We refer to this non-LSTM variant as just \textbf{GNN (G1)} and the full model described in Section \ref{ssec:graphmethods} as \textbf{GNN-LSTM (G2)}. The results for these 4 models and 5 disease phenotypes are reported in Table \ref{tab:clinical-baselines}. We observe that graph-based methods consistently outperform the baseline approaches. The GNN (G1) performs better than the equivalent MLP-based clinical baseline for family history (B2), demonstrating that the use of a GNN model in itself improves predictive performance. The GNN-LSTM model, which also incorporates detailed medical histories of individual family members, was the best performing method for all diseases except colorectal cancer. Since cancers are known to be the least hereditary out of the diseases we studied, we hypothesize that the use of GNNs for modeling family information provides the most benefit for more heritable diseases and study this further using synthetic data in Section \ref{sssec:experiments-synthetic}. 

\begin{landscape}
\begin{table}[h]
\footnotesize
\caption{Comparison of predictive performance for GNN and GNN-LSTM models with clinical baselines. Reports the average value for each metric and 95\% CI of the uncertainty estimates in brackets (except for the rule-based approach), with best results indicated in bold.}
\begin{tabular}{p{0.1\linewidth} p{0.31\linewidth} p{0.17\linewidth} p{0.17\linewidth} p{0.17\linewidth}}
  \toprule
 \textbf{Endpoint} & \textbf{Model} & \textbf{AUC-ROC} & \textbf{AUC-PRC} & \textbf{MCC} \\
   \midrule
\multirow{4}{0.1cm}{Coronary heart disease} & B1: Clinical baseline binary & 0.573 & 0.039 & 0.052 \\
 & \multirow{1}{*}{B2: Clinical baseline MLP} & 0.710 (0.7097 - 0.71031) & 0.073 (0.0721 - 0.0739) & 0.114 (0.1071 - 0.1209) \\
 & \multirow{1}{*}{G1: GNN} & 0.720 (0.7198 - 0.7202)  & 0.079 (0.0789 - 0.0791) & 0.125 (0.0590 - 0.191) \\
 & \multirow{1}{*}{G2: GNN-LSTM} & \textbf{0.775 (0.7747 - 0.7754)} & \textbf{0.126 (0.1249 - 0.1271)} & \textbf{0.169 (0.0800 - 0.2580)} \\ \hline
\multirow{4}{0.1cm}{Type 2 diabetes} & B1: Clinical baseline binary & 0.582 & 0.075 & 0.085 \\
 & \multirow{1}{*} {B2: Clinical baseline MLP} & 0.662 (0.6615 - 0.6625) & 0.111 (0.1108 - 0.1112) & 0.114 (0.0530 - 0.1750) \\
 & \multirow{1}{*}{G1: GNN} & 0.673 (0.6727 - 0.6733) & 0.117 (0.1168 - 0.1172) & 0.122 (0.0590 - 0.1850) \\
 & \multirow{1}{*}{G2: GNN-LSTM} & \textbf{0.700 (0.6994 - 0.7006)} & \textbf{0.127 (0.1263 - 0.1277)} & \textbf{0.141 (0.0700 - 0.2120)} \\ \hline
\multirow{4}{0.1cm}{Depression} & B1: Clinical baseline binary & 0.557 & 0.072 & 0.056 \\
 & \multirow{1}{*}{B2: Clinical baseline MLP} & 0.639 (0.6387 - 0.6393) & 0.105 (0.1043 - 0.1057) & 0.102 (0.0450 - 0.1590) \\
 & \multirow{1}{*}{G1: GNN} & 0.642 (0.6419 - 0.6421) & 0.112 (0.1119 - 0.1121) & 0.111 (0.0550 - 0.1670) \\
 & \multirow{1}{*}{G2: GNN-LSTM} & \textbf{0.655 (0.6548 - 0.6552)} & \textbf{0.119 (0.1183 - 0.1197)} & \textbf{0.116 (0.0570 - 0.1750)} \\ \hline
\multirow{4}{0.1cm}{Asthma} & B1: Clinical baseline binary & 0.542 & 0.013 & 0.019 \\
 & \multirow{1}{*}{B2: Clinical baseline MLP} & 0.635 (0.6347 - 0.6352) & 0.020 (0.0195 - 0.0205) & 0.031 (0.0140 - 0.0480) \\
 & \multirow{1}{*}{G1: GNN} & 0.647 (0.6467 - 0.6473) & 0.021 (0.0204 - 0.0216) & 0.039 (0.0210 - 0.0570) \\
 & \multirow{1}{*}{G2: GNN-LSTM} & \textbf{0.666 (0.6657 - 0.6661)} & \textbf{0.023 (0.0228 - 0.0232)} & \textbf{0.042 (0.0418 - 0.0422)} \\  \hline
\multirow{4}{0.1cm}{Colorectal cancer} & B1: Clinical baseline binary & 0.524 & 0.009 & 0.019 \\
 & \multirow{1}{*}{B2: Clinical baseline MLP} & 0.650 (0.6497 - 0.6503) & 0.015 (0.0146 - 0.0154) & 0.032 (0.0150 - 0.0490) \\
 & \multirow{1}{*}{G1: GNN} & \textbf{0.654 (0.6534 - 0.6546)} & \textbf{0.016 (0.0160 - 0.0160)} & \textbf{0.035 (0.0170 - 0.0530)} \\
 & \multirow{1}{*}{G2: GNN-LSTM} & 0.653 (0.6522 - 0.6538) & \textbf{0.016 (0.0158 - 0.0162)} & 0.031 (0.0200 - 0.0420) \\ \bottomrule
\end{tabular}
\label{tab:clinical-baselines}
\end{table}
\end{landscape}

\subsection{Ablation studies}
\label{ssec:ablation-fin}

The GNN-LSTM results may be of broader interest for the development of machine learning approaches for longitudinal EHR data, so we further examine this method with ablation studies. Starting from a baseline of a 2-layer \textbf{MLP for age and sex features (A1)}, we compare:

\begin{itemize}
    \item \textbf{Age, sex and family history MLP (A2, same as B2)}: adding family history features, as defined in Section \ref{ssec:features};
    \item \textbf{Age, sex and graph connectivity MLP (A3)}: adding topological information with an 8-dimensional Node2Vec embedding \citep{grover2016node2vec} for each patient derived from the connected family network for the entire dataset;
    \item \textbf{Age, sex and longitudinal EHR data LSTM (A4)}: adding longitudinal EHR data (for the target patient), using a single-layer, bidirectional LSTM architecture, i.e., similar to the target-specific component of the GNN-LSTM model. The choice of neural network architecture is based on a recent review which observes that single-layer LSTMs are frequently used for deep patient time series prediction and bidirectional variants consistently outperform unidirectional counterparts \citep{10.1145/3531326};
    \item \textbf{Age, sex, family history and longitudinal EHR data LSTM (A5)} the LSTM model (A4), but also including family history as static features. Adding family history features to an LSTM is different to modeling this with a GNN-LSTM and so this model helps us understand the benefits (if any) of using GNNs.
\end{itemize}

\noindent The results for these models compared to the \textbf{GNN-LSTM (AG, same as G2)} are shown in Table \ref{tab:ablation-fin}. Most of the gain over the age and sex baseline (A1) comes from the addition of family history (A2) and longitudinal EHR data (A4), as opposed to topological information alone (A3). When all this information is combined into a single model using the GNN-LSTM approach (AG), this generally performs better than individual approaches (A1-5), especially for highly multifactorial diseases such as coronary heart disease and type 2 diabetes. The GNN-LSTM (AG) performed better than an LSTM that includes static features for family history (A5), demonstrating that graphs are a more appropriate model for family information. However, an LSTM without family history (A4) obtained higher MCC than the GNN-LSTM (AG) for asthma and colorectal cancer, which may indicate that for some diseases a geometric prior based on genetic relatedness alone is not sufficient (e.g., because of stronger non-genetic effects). Since it is difficult to distinguish genetic and environmental effects for real EHR data we use synthetic data to study the effects of heritability in isolation in Section \ref{sssec:experiments-synthetic}.

\begin{landscape}
\begin{table}[]
\footnotesize
\caption{Ablation studies examining how different aspects modeled by the GNN-LSTM contribute to the performance: covariates age and sex (A1), family history (A2), family relatedness (A3), and detailed medical histories (A4). A GNN-LSTM (AG) is a graph-based model for this information, which is a conceptually different approach to adding features for family history to an LSTM model (A5). Reports the average value for each metric and 95\% CI of the uncertainty estimates in brackets, with best results indicated in bold.}
\begin{tabular}{p{0.1\linewidth} p{0.31\linewidth} p{0.17\linewidth} p{0.17\linewidth} p{0.17\linewidth}}
  \toprule
 \textbf{Endpoint} & \textbf{Model} & \textbf{AUC-ROC} & \textbf{AUC-PRC} & \textbf{MCC} \\ 
 \midrule
\multirow{4}{0.1cm}{Coronary heart disease} & A1: Age and sex (MLP) & 0.696 (0.6956 - 0.6964) & 0.066 (0.0656 - 0.0665) & 0.093 (0.0820 - 0.1040) \\
 & A2: Age, sex, family history (MLP) & 0.710 (0.7097 - 0.7103) & 0.073 (0.0721 - 0.0739) & 0.114 (0.1071 - 0.1209) \\
 & A3: Age, sex, graph connectivity (MLP) & 0.696 (0.6956 - 0.6964) & 0.067 (0.0669 - 0.0671) & 0.091 (0.0818 - 0.1002) \\
 & A4: Age, sex, EHR (LSTM) & 0.763 (0.7625 - 0.7635) & 0.105 (0.1044 - 0.1056) & 0.148 (0.1250 - 0.1710) \\
 & A5: Age, sex, EHR, family history (LSTM) & 0.771 (0.7706 - 0.7714) & 0.123 (0.1228 - 0.1232) & 0.164 (0.1580 - 0.1700) \\
 & AG: GNN-LSTM & \textbf{0.775 (0.7746 - 0.7754)} & \textbf{0.126 (0.1249 - 0.1271)} & \textbf{0.169 (0.0800 - 0.2580)} \\ \hline
\multirow{4}{0.1cm}{Type 2 diabetes} & A1: Age and sex (MLP) & 0.617 (0.6166 - 0.6174) & 0.085 (0.0848 - 0.0853) & 0.072 (0.0330 - 0.1110) \\
 & A2: Age, sex, family history (MLP) & 0.662 (0.6615 - 0.6625) & 0.111 (0.1108 - 0.1112) & 0.114 (0.0530 - 0.1750) \\
 & A3: Age, sex, graph connectivity (MLP) & 0.619 (0.6187 - 0.6193) & 0.087 (0.0870 - 0.0870) & 0.073 (0.0340 - 0.1120) \\
& A4: Age, sex, EHR (LSTM) & 0.675 (0.6745 - 0.6755) & 0.110 (0.1091 - 0.1109) & 0.111 (0.0520 - 0.1700) \\
 & A5: Age, sex, EHR, family history (LSTM) & 0.689 (0.6884 - 0.6896) & 0.121 (0.1200 - 0.1220) & 0.127 (0.0600 - 0.1940) \\
 & AG: GNN-LSTM & \textbf{0.700 (0.6994 - 0.7006)} & \textbf{0.127 (0.1263 - 0.1277)} & \textbf{0.141 (0.0700 - 0.2120)} \\ \hline
\multirow{4}{0.1cm}{Depression} & A1: Age and sex (MLP) & 0.632 (0.6318 - 0.6323) & 0.102 (0.1018 - 0.1022) & 0.097 (0.0450 - 0.1490) \\
 & A2: Age, sex, family history (MLP) & 0.639 (0.6387 - 0.6393) & 0.105 (0.1043 - 0.1057) & 0.102 (0.0450 - 0.1590) \\
 & A3: Age, sex, graph connectivity (MLP) & 0.633 (0.6328 - 0.6332) & 0.102 (0.1017 - 0.1023) & 0.098 (0.0460 - 0.1500) \\
& A4: Age, sex, EHR (LSTM) & 0.647 (0.6467 - 0.6473) & 0.113 (0.1121 - 0.1139) & 0.108 (0.0510 - 0.1650) \\
 & A5: Age, sex, EHR, family history (LSTM) & 0.651 (0.6508 - 0.6512) & \textbf{0.119 (0.1183 - 0.1197)} & 0.111 (0.0520 - 0.1700) \\
& AG: GNN-LSTM & \textbf{0.655 (0.6548 - 0.6552)} & \textbf{0.119 (0.1183 - 0.1197)} & \textbf{0.116 (0.0570 - 0.1750)} \\ \hline
\multirow{4}{0.1cm}{Asthma} & A1: Age and sex (MLP) & 0.625 (0.6249 - 0.6251) & 0.018 (0.0179 - 0.0181) & 0.034 (0.0160 - 0.0520) \\
 & A2: Age, sex, family history (MLP) & 0.635 (0.6348 - 0.6352) & 0.020 (0.0195 - 0.0205) & 0.031 (0.0140 - 0.0480) \\
 & A3: Age, sex, graph connectivity (MLP) & 0.622 (0.6217 - 0.6223) & 0.017 (0.0168 - 0.0172) & 0.034 (0.0160 - 0.0520) \\
& A4: Age, sex, EHR (LSTM) & 0.653 (0.6526 - 0.6534) & 0.021 (0.0209 - 0.0211) & \textbf{0.043 (0.0200 - 0.0660)} \\
 & A5: Age, sex, EHR, family history (LSTM) & 0.664 (0.6636 - 0.6645) & 0.022 (0.0219 - 0.0222) & 0.037 (0.0170 - 0.0570) \\
& AG: GNN-LSTM & \textbf{0.666 (0.6659 - 0.6661)} & \textbf{0.023 (0.0228 - 0.0232)} & 0.042 (0.0418 - 0.0422) \\ \hline
\multirow{4}{0.1cm}{Colorectal cancer} & A1: Age and sex (MLP) & 0.651 (0.6505 - 0.6515) & 0.015 (0.0149 - 0.0151) & 0.026 (0.0120 - 0.0400) \\
 & A2: Age, sex, family history (MLP) & 0.650 (0.6497 - 0.6503) & 0.015 (0.0146 - 0.0154) & 0.032 (0.0150 - 0.0490) \\
 & A3: Age, sex, graph connectivity (MLP) & 0.645 (0.6447 - 0.6453) & 0.015 (0.0150 - 0.0150) & 0.034 (0.0160 - 0.0520) \\
& A4: Age, sex, EHR (LSTM) & 0.652 (0.6519 - 0.6521) & 0.015 (0.0149 - 0.0152) & \textbf{0.035 (0.0160 - 0.0540)}  \\
 & A5: Age, sex, EHR, family history (LSTM) & 0.646 (0.6456 - 0.6464) & 0.015 (0.0149 - 0.0151) & 0.027 (0.0130 - 0.0410) \\
 & AG: GNN-LSTM & \textbf{0.653 (0.6522 - 0.6538)} & \textbf{0.016 (0.0158 - 0.0162)} & 0.031 (0.0200 - 0.0420) \\
 \bottomrule
\end{tabular}
\label{tab:ablation-fin}
\end{table}
\end{landscape}

\subsection{Synthetic data experiments}
\label{sssec:experiments-synthetic}

\begin{figure}[t]
  \centering 
  \includegraphics[width=\textwidth]{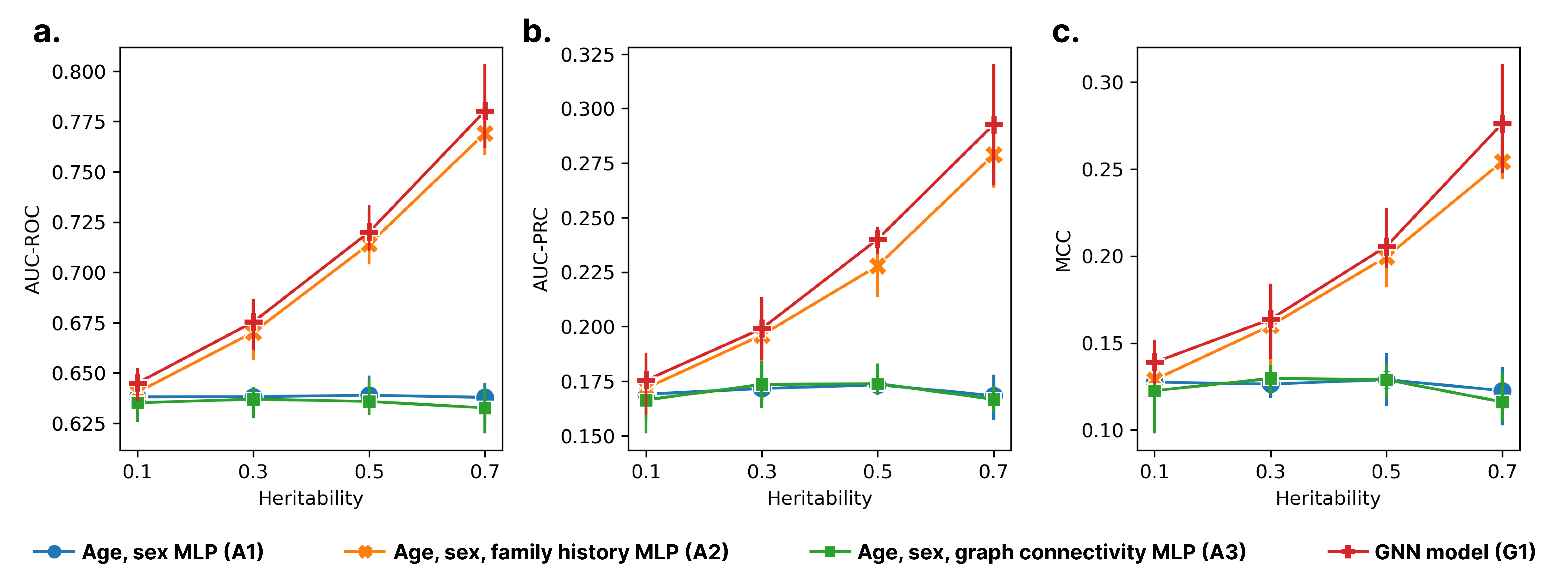} 
  \caption{Ablation study for synthetic data with phenotypes simulated with varying degrees of heritability, $h$. Reports (a) AUC-ROC, (b) AUC-PRC and (c) MCC averaged across 5 simulations of each dataset $h=0.1, 0.3, 0.5, 0.7$, plotted with error bars.}
  \label{fig:heritability} 
\end{figure}

We use synthetic datasets generated by the algorithm described in Section \ref{ssec:cohort-synthetic} to examine phenotypes of differing degrees of heritability, $h$. Results for an ablation study (Figure \ref{fig:heritability}) show that for various simulated phenotypes $h=0.1, 0.3, 0.5, 0.7$, the GNN method (G1) consistently outperforms the baselines and this performance gain improves for more heritable phenotypes. The result that the most gain comes from the addition of family information (A2), but modeling this as a geometric prior using a GNN (G1) provides further gains, is consistent with what was observed for real data (Table \ref{tab:ablation-fin}). We also use synthetic data to examine the sensitivity of the GNN model to various design choices (Figure \ref{fig:gnn-design-syn}). These results show marginal differences in AUC-ROC for the design choices considered. The standard approach for constructing a family tree uses parent/child edges, but we found that better performance was achieved for GNNs by also adding direct edges from each node to the target node, or edges for all genetically related nodes. The average improvement in AUC-ROC was 0.22\% and 0.28\%, respectively, but varied depending on the choice of GNN architecture. Overall, the highest AUC-ROC was attained with k-GNN layers \citep{morris2019weisfeiler} and target node representations (no pooling), for family graphs with edges connecting all genetically related nodes.

\subsection{Explainability analysis}

\begin{figure}[h]
  \centering 
  \includegraphics[width=\textwidth]{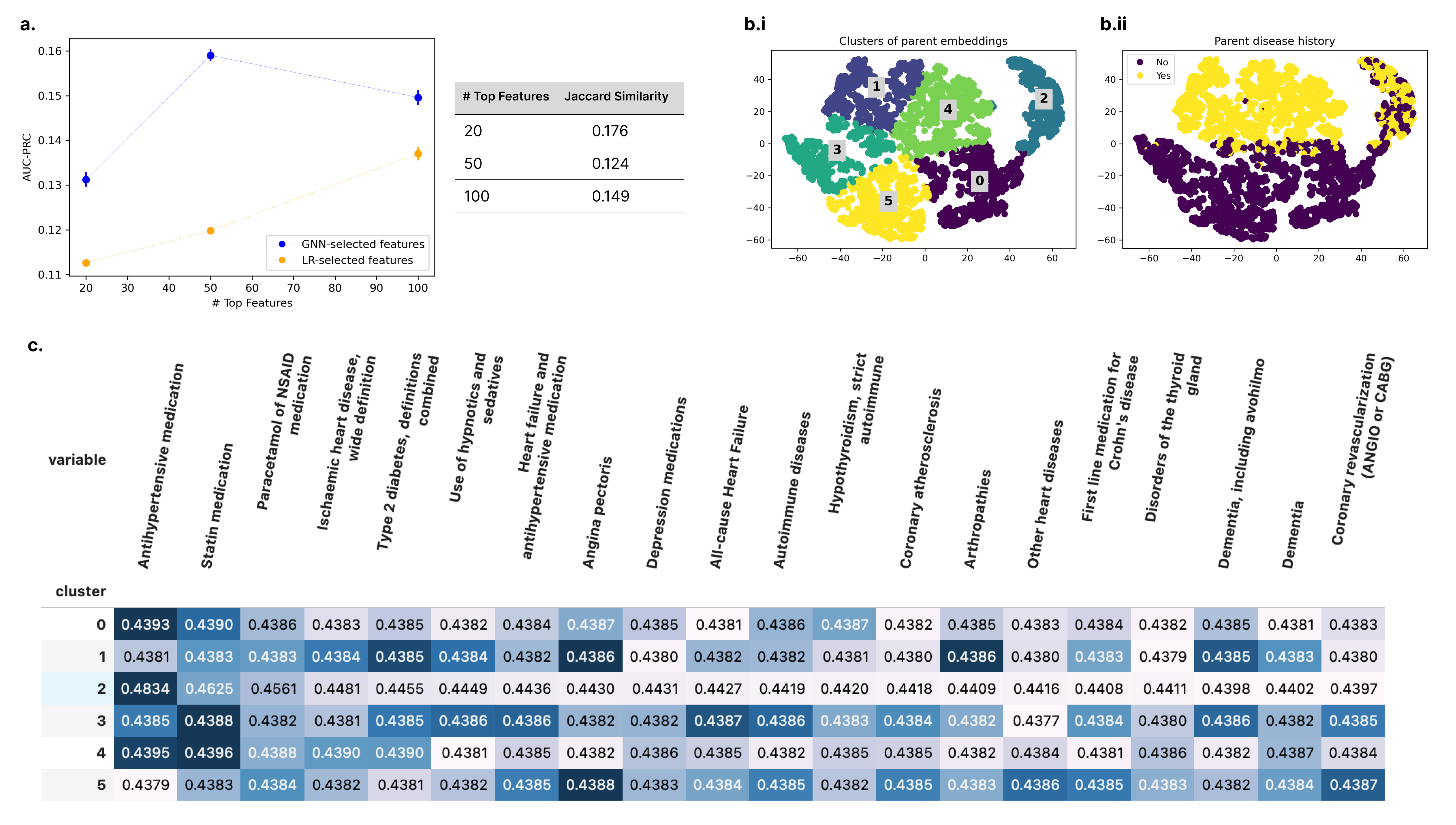} 
  \caption{GNN-LSTM explanations for parent node embeddings, CHD endpoint. \textbf{a)} AUC-PRC is plotted for GNNs retrained with the top-$n$ features identified by (1) the GNN-LSTM and (2) a logistic regression (LR) baseline. The Jaccard similarity between the GNN- and LR-selected features is also reported; \textbf{b.i)} Graph representations of parents visualized using t-SNE and k-means clustering; \textbf{b.ii)} The representations clearly identify family history as a determining factor for the child's risk; \textbf{c)} Feature importances for each cluster, for the top-20 features identified by the model overall.}
  \label{fig:exp-eval} 
\end{figure}

We use the GNN-LSTM model for coronary heart disease (CHD) as an illustrative example for analyzing the graph representations learned by the model, using the graph explainability methods described in Section \ref{sssec:explainability-methods}. Since the ground truth important features for each relative are not available for the entire EHR feature set, we compare the GNN-LSTM approach with a logistic regression baseline - a commonly used method in epidemiological studies. We retrain GNNs\footnote{For the GNN model, the longitudinal EHR features are aggregated as 1 if the diagnosis was recorded at least once during the observation period and 0 otherwise} with the top-$n$ EHR features identified for parents by (1) GNN-LSTM explainability and (2) a logistic regression baseline, for $n=20,50,100$. We observe a higher AUC-PRC (19.5\% gain on average) for features identified by the GNN-LSTM model (Figure \ref{fig:exp-eval}a), indicating that the graph-based approach is better at identifying features for parents that are useful for predicting a child's CHD risk. 

As a case study, we further examine the node embeddings learned by the GNN-LSTM for the parent nodes of CHD patients. The embeddings are visualized using the t-SNE algorithm and k-means clustering is used to identify groups of similar parents (Figure \ref{fig:exp-eval}b.i), with the corresponding feature importances for each cluster shown in Figure \ref{fig:exp-eval}c. We observe well established CHD risk factors, comorbidities and treatments such as antihypertensive medication, statins, T2D and various heart diseases \citep{escobar2002hypertension,goodarzi2020genetics,mega2015genetic} from the parent being important predictors of 10 year CHD outcomes in their offspring. Figure \ref{fig:exp-eval}b.ii shows that the embeddings learn to separate a parent's predictive influence based on their CHD history, with certain clusters being more associated with a parent having CHD than others. However, while this explainability analysis is beneficial for identifying features that the GNN-LSTM finds useful for predicting disease, we note that it does not directly provide a causal interpretation of a feature’s effect on disease.

\section{Discussion and conclusion} 

In this work we developed a new geometric deep learning approach to model the relational structure of high-dimensional, longitudinal medical histories within families and learn explainable, supervised graph representations of a patient's disease risk. We compared this approach with (statistical) clinical baselines for family history and deep learning baselines for EHR data and showed that graph-based approaches achieved higher AUC-ROC, AUC-PRC and MCC for predicting a patient’s 10-year disease onset for 5 complex disease phenotypes (Tables \ref{tab:clinical-baselines} and \ref{tab:ablation-fin}). Ablation studies for synthetic data indicated that a GNN-based approach provided the most gains in predictive performance for more heritable diseases (Figure \ref{fig:heritability}) and ablation studies for real data demonstrated that the GNN itself contributed to these gains, i.e., compared to equivalent non-graph MLP and LSTM baselines for family history (Tables \ref{tab:clinical-baselines} and \ref{tab:ablation-fin}). These results illustrate that the underlying geometric structure of family history is suitably modeled using graph-based approaches. Furthermore, using explainability techniques, we showed how a graph-based approach allows us to characterize the influences of family information in a personalized manner by identifying the individual relatives and features that were most important for the model’s prediction of a patient’s disease risk. We observed that GNNs gained 19.5\% AUC-PRC on average when retrained with important features identified by the GNN-LSTM model, compared to features identified using a logistic regression baseline. 

\subsection{Limitations}

Limitations of our analysis included a lack of comparisons to disease-specific methods for interpreting family history, as we aimed to provide a broader study of a disease-agnostic approach. We acknowledge that any new machine learning approach requires further development to operationalize in a clinical setting and that ethical and legal concerns may impact the availability of family EHR data. In the case where there is restricted access to family EHRs, we propose to first train the model on a large scale EHR dataset and use the graph explainability analysis to identify important features. In a clinical setting, if family EHR data is available  the model can provide personalized inference, otherwise population averages from the training data can be used to provide general guidance on what aspects of family history (up to third-degree relatives) to discuss with a patient.

Our work focused on examining whether geometric approaches can be beneficial for modeling family history. We limited the analysis to existing GNN architectures as our study is a first step in evaluating the suitability of geometric approaches for family history. Future research could develop specialized GNN layers for this machine learning problem. Class imbalance and intra-class variability of heterogeneous diseases posed machine learning challenges, which we alleviated using techniques described in Appendix \ref{app:class-imbalance}. We also took measures to reduce potential confounding factors in our cohort design and machine learning problem setup, but this means that some aspects of the real data are not explicitly modeled in our analysis, e.g., uncertainty in missing family information, children nodes in family graphs. We also noted that while our explainability analysis identifies nodes and features that are useful for prediction, this does not directly translate to causal interpretations (e.g., importance may be due to correlation).

\subsection{Conclusion}

Overall, we believe that the machine learning approaches developed in our work present new opportunities for clinical, epidemiological, and precision medicine applications. We demonstrate how a graph-based approach provides more informative representations for predicting disease risk incorporating family history. A GNN-LSTM that models family EHRs as a connected network learns explainable node embeddings for individual relatives and achieves better predictive performance for complex diseases than clinically-inspired and deep learning baselines. Future work in this direction can benefit clinical applications such as personalized family history assessments, as well as epidemiological studies of how familial relationships affect health and disease.

\acks{We thank the FinRegistry team for making the data available for this study, and we acknowledge CSC – IT Center for Science, Finland, for computational resources. This study has received funding from the European Union’s Horizon 2020 research and innovation programme under grant agreement No 101016775. FinRegistry is a collaboration project of the Finnish Institute for Health and Welfare (THL) and the Data Science Genetic Epidemiology research group at the Institute for Molecular Medicine Finland (FIMM), University of Helsinki. The FinRegistry project has received the following approvals for data access from the National Institute of Health and Welfare (THL/1776/6.02.00/2019 and subsequent amendments), DVV (VRK/5722/2019-2), Finnish Center for Pension (ETK/SUTI 22003) and Statistics Finland (TK-53-1451-19). The FinRegistry project has received IRB approval from the National Institute of Health and Welfare (Kokous 7/2019).}

\bibliography{references}

\appendix
\section{Model training details}
\label{appendix:hyperparameter-choices-etc}

\begin{figure}[h]
  \centering 
  \includegraphics[width=\textwidth]{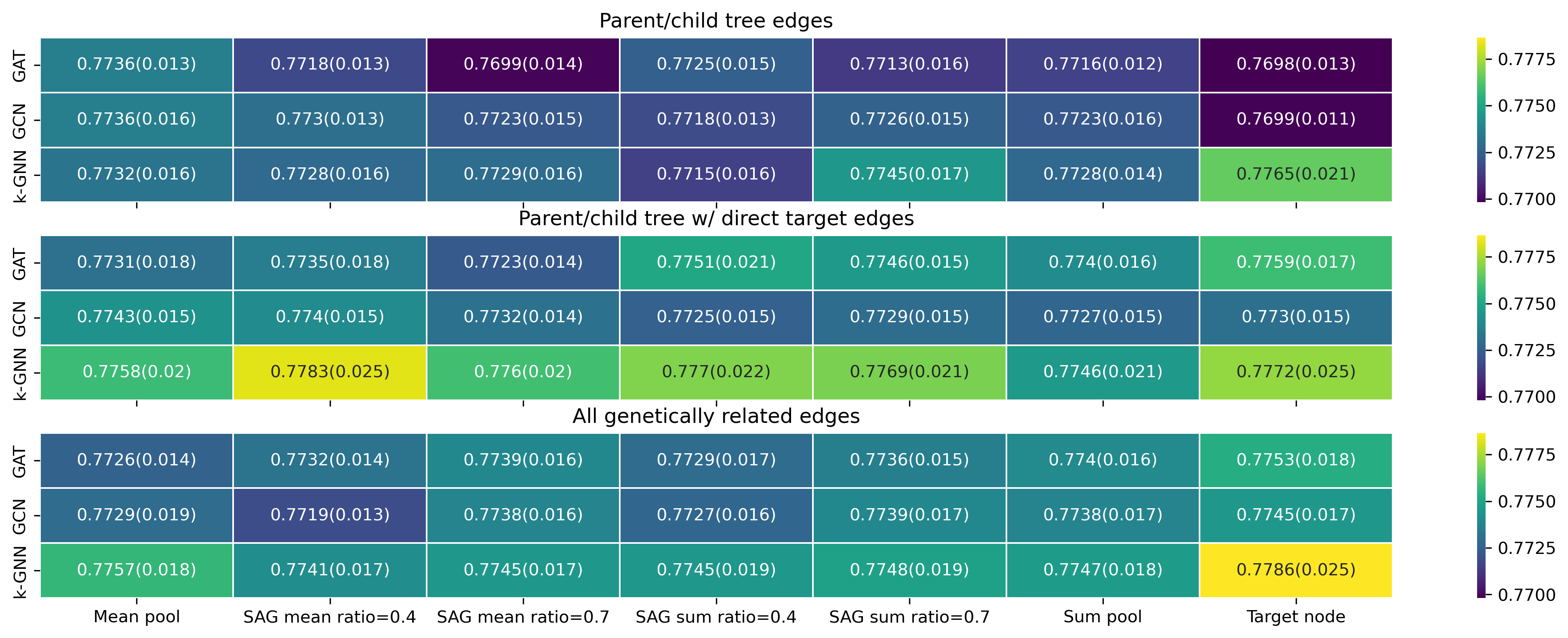} 
  \caption{GNN design study, reporting AUC-ROC averaged across 5 trials (and 95\% confidence intervals in brackets) of $h=0.7$ synthetic datasets for hyperparameters listed in Table \ref{tab:hyperparameter-values}.}
  \label{fig:gnn-design-syn} 
\end{figure}

Table \ref{tab:hyperparameter-values} lists hyperparameter values and architecture design choices used in experiments. For synthetic data experiments, a grid search was performed on all combinations listed, with results shown in Figure \ref{fig:gnn-design-syn}. For real data experiments, due to computational constraints in the sensitive data computing environment, we used the values corresponding to the best-performing architecture identified in the synthetic data experiment. The baseline methods compared in experiments use the same hyperparameter values, where applicable. Computational complexity is reported in Table \ref{tab:computational-time}.

\begin{table}[h]
  \centering 
  \footnotesize
  \caption{Hyperparameter values and architecture design choices. Values used for the real data experiments are indicated in bold.}
  \begin{tabular}{p{0.35\linewidth} | p{0.6\linewidth}}
  \toprule
    \textbf{Description} & \textbf{Values} \\ 
    \midrule
Edge definitions & Parent/child, parent/child with target, \textbf{all genetically related} \\
GNN layer & GCN \citep{kipf2016semi}, \textbf{k-GNN} \citep{morris2019weisfeiler}, GAT \citep{velivckovic2017graph} \\
Pooling layer & \textbf{Target}, Sum, Mean, SAGpool (sum), SAGpool (mean) \citep{lee2019self, liu2022graph} \\
SAGpool ratio & 0.4, 0.7 \\
Optimizer & Adam with $\beta_1=0.9, \beta_2=0.999$ \\
Batchsize & 250 \\
Max epochs & 100 \\
Patience (early stopping criteria) & 5 \\
Learning rate & 0.01, \textbf{0.001} \\
Hidden dimensions - GNN units & $H_G = 20$ \\
Hidden dimensions - LSTM units & $H_L = 40$ \\
Hidden dimensions - MLP units & $H_M = 20$ \\
$\gamma$ & 1 \\
$\alpha$ & 1 \\
$\beta$ & 1 \\
$\delta$ & 1 \\
Dropout rate & 0.5 \\
    \bottomrule
  \end{tabular}
  \label{tab:hyperparameter-values} 
\end{table}

\subsection{Methods for managing class imbalance during training}
\label{app:class-imbalance}

\begin{itemize}
    \item Sampling strategies: The train/validation/test split is setup by splitting the minority class (cases) into train, validation and test sets with a 70:10:20 ratio and randomly allocating controls to these sets by undersampling the majority class (controls) with a 15:85 case-control ratio in the training set. This improves statistical power during training, but the dataset case-control ratio is used in the validation and test sets to maintain a realistic population distribution for validation and testing.
    \item Class-weighted loss: Loss-based approaches have been shown to aid imbalanced classification problems in the health domain \citep{depto2023quantifying}. We weight the BCE loss terms by class ratios, to more highly penalize misclassification of the minority class. 
    \item Overfitting strategies: To prevent overfitting the minority class, dropout layers are used and an early stopping criterion is calculated on the validation set, to stop training if the validation losses start increasing. The classification threshold is regarded as a hyperparameter and is optimized by applying threshold-moving methods on the validation set \citep{krawczyk2015cost, zhou2005training} based on the precision-recall curve.
    \item Evaluation metrics: Class imbalance can bias the interpretation of widely used metrics such as AUC-ROC \citep{hicks2022evaluation, saito2015precision}. A variety of metrics are used for evaluating performance on the test set: AUC-ROC, AUC-PRC and Matthews correlation coefficient (MCC).
\end{itemize}

\begin{table}[h]
  \centering 
  \footnotesize
  \caption{Average training time (and standard deviation) in minutes for the 5 diseases. Models were trained on a single NVIDIA A100 GPU.}
  \begin{tabular}{ll}
  \toprule
    \textbf{Model} & \textbf{Average training time} \\
    \midrule
        Age, sex MLP (A1) & 12.98 (5.07) \\ 
        Age, sex, family history MLP (A2) & 19.30 (7.12) \\ 
        Age, sex, graph connectivity (A3) & 15.56 (7.90) \\ 
        GNN (G1) & 59.88 (28.20) \\ 
        Age, sex, EHR LSTM (A4) & 88.92 (26.66) \\ 
        Age, sex, EHR LSTM, family history (A5) &  84.34 (26.37) \\ 
        GNN-LSTM (G2) & 381.90 (165.73) \\
    \bottomrule
  \end{tabular}
  \label{tab:computational-time} 
\end{table}

\section{Synthetic data algorithm} \label{AppxC}
Under assumptions mentioned in Section \ref{ssec:cohort-synthetic}, we generate synthetic data using the algorithm as below: \\

\begin{enumerate}
    \item Generate individual $i$ for ancestry generation (generation 0) with \\
$$L^0_{Herr,i} \sim N(0, h^2), \epsilon^0_{i} \sim N(0, e^2), Gender^0_{i} \sim \mathrm{Ber}(0.5), Age^0_{i} \sim \mathrm{Uni}(0, 1),$$
and his total phenotypic liability $ L^0_{i} = L^0_{Herr,i} + \beta_{age} \cdot Age^0_{i} + \beta_{sex} \cdot Gender^0_{i} + \epsilon^0_{i}$. $h^2, e^2, \beta_{age}$ and $\beta_{sex}$ are all input parameters to adjust for contribution of each component on the total liability.
    \item Individual $j$ from generation $k (k>0)$ can then be generated as below\\
$$\epsilon^k_{j} \sim N(0, e^2), Gender^k_{j} \sim \mathrm{Ber}(0.5), Age^k_{j} \sim \mathrm{Uni}(0, 1),$$
 and $L^k_{Herr,j}= c \cdot (\alpha_1 L^{k-1}_{Herr,mo} + \alpha_2 L^{k-1}_{Herr,fa}) $
where $mo, fa$ are randomly selected mother and father from the previous generation that satisfy:
    \begin{enumerate}
        \item Mother should be female and father should be male;
        \item Mother and father are not related within three generations (they are not the sample person, don’t share parents, or grandparents)
    \end{enumerate}
\end{enumerate}
$\alpha_1$ and $\alpha_2$ are parameters for parental contribution on liability of the offspring and $\alpha_1 + \alpha_2 = 1$. Since by definition $L_{Herr}$ covers both genetic heritable risk and non-generic parental impact, they are not necessarily equal. $c = \frac{1}{\alpha_1^2 + \alpha_2}$ is a constant to stabilize liability variance in future generations. Total phenotypic liability for this individual will be $ L^k_{j} = L^k_{Herr,j} + \beta_{age} \cdot Age^k_{j} + \beta_{sex} \cdot Gender^k_{j} + \epsilon^k_{j}$. We repeat step 2 until the designated number of generations are reached. To map continuous phenotypic liability onto binary disease status, we can assume a disease prevalence $k$ and set a cutoff threshold $\Phi^{-1}(k)$ on the standardized liability. We then assign all $L^k_{i} \geq \Phi^{-1}(k)$ as cases and otherwise controls.

\section{Endpoint definitions}

\begin{table}[h]
\caption{Endpoint details. Disease specific age ranges and control exclusion criteria}
\footnotesize
\begin{tabular}{p{0.3\linewidth}  p{0.2\linewidth}  p{0.5\linewidth}}
\toprule
 \textbf{Disease of interest
} & \textbf{Target age range} & \textbf{Target exclude conditions}  \\ 
   \midrule
Adult asthma & 20-70 & Any asthma, chronic obstructive pulmonary disease (COPD) and relevant medical reimbursement \\
Coronary heart disease (CHD) & 40-70 & Any cardiovascular conditions \\
Colorectal cancer & 45-75 & Any cancer diagnosis \\
Depression & 18-60 & Any mood disorders, suicide attempt, anti-depressant purchase \\
Type ii diabetes (T2D) & 35-70 & Type i,ii diabetes and relevant complications, insulin purchase \\
\bottomrule
\end{tabular}
\label{tab:endpoint-definitions}
\end{table}

\end{document}